\documentclass[twocolumn,3p]{elsarticle}

\usepackage{amsmath,amssymb}
\usepackage{natbib}
\usepackage{graphicx}
\usepackage{dcolumn}
\usepackage{bm}

\usepackage[breaklinks=true,pdffitwindow=false,bookmarks=true,pdfauthor={Hoffmann et al.},colorlinks=true,citecolor=blue,linkcolor=blue]{hyperref}
\usepackage{breakurl}
\usepackage{hypernat}
\usepackage{hypcap}

\makeatletter
\newcommand*{\defeq}{\mathrel{\rlap{%
                     \raisebox{0.3ex}{$\m@th\cdot$}}%
                     \raisebox{-0.3ex}{$\m@th\cdot$}}%
                     =}
\makeatother

\begin{document}

\title{{\tt kmos}: A lattice kinetic Monte Carlo framework}

\author[tum]{Max J. Hoffmann\corref{cor1}}
\ead{max.hoffmann@ch.tum.de}
\author[tum]{Sebastian Matera}
\author[tum]{Karsten Reuter}
\ead{karsten.reuter@ch.tum.de}
\address[tum]{Chair of Theoretical Chemistry and Catalysis Research Center,\\ Technische Universit{\"a}t M{\"u}nchen, Lichtenbergstr. 4, D-85747 Garching, Germany}

\cortext[cor1]{Corresponding Author}

\begin{abstract}
Kinetic Monte Carlo (kMC) simulations have emerged as a key tool for microkinetic modeling in heterogeneous catalysis and other materials applications. Systems, where site-specificity of all elementary reactions allows a mapping onto a lattice of discrete active sites, can be addressed within the particularly efficient lattice kMC approach. To this end we describe the versatile {\tt kmos} software package, which offers a most user-friendly implementation, execution, and evaluation of lattice kMC models of arbitrary complexity in one- to three-dimensional lattice systems, involving multiple active sites in periodic or aperiodic arrangements, as well as site-resolved pairwise and higher-order lateral interactions. Conceptually, {\tt kmos} achieves a maximum runtime performance which is essentially independent of lattice size by generating code for the efficiency-determining local update of available events that is optimized for a defined kMC model. For this model definition and the control of all runtime and evaluation aspects {\tt kmos} offers a high-level application programming interface. Usage proceeds interactively, via scripts, or a graphical user interface, which visualizes the model geometry, the lattice occupations and rates of selected elementary reactions, while allowing on-the-fly changes of simulation parameters. We demonstrate the performance and scaling of {\tt kmos} with the application to kMC models for surface catalytic processes, where for given operation conditions (temperature and partial pressures of all reactants) central simulation outcomes are catalytic activity and selectivities, surface composition, and mechanistic insight into the occurrence of individual elementary processes in the reaction network. 
\end{abstract}


\maketitle

\noindent
{\bf PROGRAM SUMMARY}\\
\begin{small}
{\em Manuscript Title: {\tt kmos}: A lattice kinetic Monte Carlo framework}     \\
{\em Authors: MJ Hoffmann, S Matera, K Reuter}               \\
{\em Program Title: {\tt kmos}}                                     \\
{\em Journal Reference:}                                      \\
{\em Catalogue identifier:}                                   \\
{\em Licensing provisions: GPLv3}                             \\
{\em Programming language: Python 16.4\%, fortran90: 83.6 \%}  \\
{\em Computer: PC, Mac}\\
{\em Operating system: Linux, Mac, Windows} \\
{\em RAM: 100MB+} \\
{\em Number of processors used: 1}                              
{\em Classification: 7 Condensed Matter and Surface Science}   \\
{\em External routines/libraries: ASE, Numpy, f2py, python-lxml} \\
{\em Subprograms used: Standard-complying Fortran compiler}     \\
{\em Nature of problem: Microkinetic simulations of complex reaction networks with all elementary processes occurring at active sites of a static lattice} \\
{\em Solution method: Efficient lattice kinetic Monte Carlo solution of the Markovian master equation underlying the reaction network}\\
{\em Restrictions: None}\\
{\em Unusual features: The framework implements a Fortran90 code generator}\\
{\em Running time: from 10 seconds to 10 hours}\\
   \\
\end{small}

\tableofcontents

\section{Introduction}

The pressing demands for ever more energy- and resource-efficient processing reinforce the long-standing quest towards a detailed mechanistic understanding of heterogeneous catalysis. At best down to the atomic level, such understanding would pave the way for a rational design of improved catalysts, which ultimately will be tailored to the nanoscale. Quantitative theory increasingly contributes to this quest with refined kinetic models that meanwhile allow to accurately predict the activity of model catalysts of increasing complexity (from single-crystal surfaces up to nanoparticles at planar supports) without any recourse to experimental data.
\cite{hansen_first-principles-based_2000,hansen_first-principles-based_2000-1,reuter_steady_2004,honkala_ammonia_2005,saeys_ab_2005,inderwildi_fischertropsch_2008,gokhale_mechanism_2008,van_santen_mechanism_2013} Such models have to span a range of scales in length and time, starting with the making and breaking of the individual chemical bonds at the electronic structure level, over the mesoscopic interplay of the various elementary reactions in the reaction network, to the heat and mass transport at the macroscopic (reactor) scale.
\cite{sabbe_first-principles_2012,salciccioli_review_2011,norskov_density_2011,honkala_expanding_????,neurock_engineering_2010,sautet_catalysis_2010,keil_multiscale_2012}

To achieve this, state-of-the-art multi-scale models resort to a hierarchical combination of different methodology. The current framework for the mesoscopic level are microkinetic approaches evaluating a (Markovian) master equation ({\em vide infra}).
\cite{sabbe_first-principles_2012,gardiner_handbook_2004,chorkendorff_concepts_2006} Using as input kinetic parameters for all elementary reactions (e.g. provided from first-principles electronic structure theory calculations), such microkinetic models determine for given operation conditions at the surface (e.g. temperature $T$ and partial pressures $\{p_{i}\}$ of all reactants $i$) not only the catalytic activity (typically measured as turn-over frequency, TOF, in units of products per active site and time) but also other important information such as surface composition, the occurrence of individual reaction steps in the network, or in particular the presence of a dominant reaction mechanism as well as rate-determining steps therein.\cite{meskine_examination_2009,temel_does_2007} Averaged over a sufficiently large catalyst surface area the TOF output can then for example be used as input for macroscale simulations of heat and mass transport in a given reactor geometry.\cite{vlachos_multiscale_1997,kissel-osterrieder_dynamic_2000,majumder_multiscale_2006,matera_first-principles_2009,matera_transport_2010,mei_effects_2011,matera_when_2012,schaefer_coupling_2013}

The traditional and still prevalent microkinetic approach employs a mean-field approximation to solve the master equation, and then only accounts for average surface coverages of the different reaction intermediates at the active surface. In case of heterogeneous arrangement of active sites, strong lateral interactions among the adsorbed species, or diffusion limitations, this approximation is known to break down and lead to qualitatively wrong results \cite{temel_does_2007,matera_adlayer_2011,wu_accurate_2012}. This has contributed to the recent rise of alternative kinetic Monte Carlo (kMC) simulations, which do not need to rely on the mean-field approximation and therefore provide a faithful account of the detailed spatial distributions of species at the catalyst surface, as well as their correlations and fluctuations.\cite{chatterjee_overview_2007,voter_introduction_2007,reuter_first-principles_2012} In contrast to effective rate equation based models for which a definition of some abstract active site (type) is often sufficient, kMC thus needs as input detailed information about the microscopic arrangement of the true active sites of the crystal surface. In return, it then provides comprehensive information about the (time-resolved) arrangement of chemicals at all these active sites during catalyst operation. Apart from a wealth of mechanistic information, e.g. about correlations in the occupation of neighboring sites at the surface, this allows to obtain proper (not erroneous mean-field) mesoscopic averages of quantities like TOFs that ultimately are required for reactor level modeling.

Due to the inherent methodological simplicity of kMC, seminal works in the surface catalysis context typically relied on specialized code written from scratch.\cite{hansen_first-principles-based_2000,hansen_first-principles-based_2000-1,reuter_steady_2004,reuter_first-principles_2006,rogal_first-principles_2007,rogal_co_2008} Even though kMC models are used in the field with increasing frequency this practice has largely prevailed and only few general kMC packages have been put forward to date.\cite{dooling_generic_2001,kunz_entwicklung_2007,garcia_cardona_parallel_2011,stamatakis_graph-theoretical_2011} This stands in stark contrast to the manifold of established and powerful software packages employed in the multi-scale framework for either the underlying electronic structure calculations \cite{_list_2013} or the continuum mechanics reactor scale simulations\cite{cfd-online_codes_2013}. Noting this as an obstacle to a further, wide-spread use of the kMC approach to surface catalysis has been the motivation for the here presented {\tt kmos} package, which as its core objective aims at a most user-friendly and efficient implementation, execution, and evaluation of increasingly complex lattice kMC models in the surface catalysis context.

\section{Theoretical Background}

\subsection{(Surface) Chemistry on a Lattice}
\label{kmc_context}

\begin{figure}
\centering{\includegraphics[width=7cm]{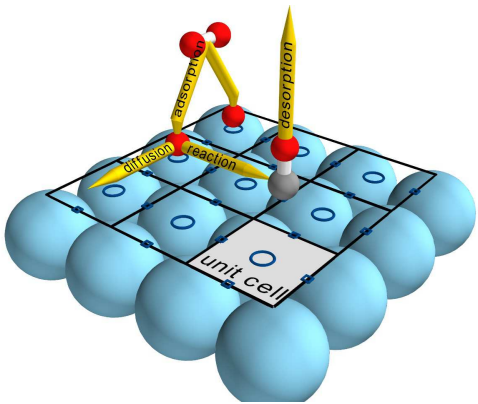}
\caption{\label{bs} (Color online) Schematic representation of a typical model catalyst surface, showing the top layer atoms of a metal(100) facet as large spheres. The periodic surface consists of repeating unit cells, each containing one or more active sites (here indicated by circles and squares for hollow and bridge adsorption sites, respectively). Possible elementary reactions in the context of CO oxidation (dissociative adsorption, diffusion, reaction and desorption) are indicated by yellow arrows.}}
\end{figure}

In terms of microkinetic modeling, the atomistic evolution proceeding during surface catalytic reactions is quite representative for a wider class of problems including crystal growth, initial corrosion, diffusion in crystalline (battery) materials or surface self-assembly. These problems feature a range of common characteristics, which motivate a so-called lattice approach to kMC that also underlies the {\tt kmos} package. In the following we use a survey over these characteristics to briefly introduce this lattice approach to kMC and clearly define terminology henceforth employed. For a more detailed account of general kMC methodology we refer to existing reviews.\cite{chatterjee_overview_2007,voter_introduction_2007,reuter_first-principles_2012} Even though the following introduction is done within the surface catalysis context, the generalization to the other problems mentioned is self-evident.

\paragraph{Site-specific adsorption and lattice mapping}
The first defining characteristic is that the surface adsorption of all reactants and reaction intermediates is assumed to be site-specific, i.e. it always occurs in well-defined so-called active {\em sites} offered by the crystalline surface. Due to the periodicity of the latter this generally leads to a lattice with each lattice point representing one such site.\cite{ashcroft_solid_1976} The actual kMC simulations only consider this lattice, which allows to encompass a wide range of system geometries within this framework. Most straightforward are extended low-index single-crystal surfaces, where the lattice is simply composed of multiple identical surface unit-cells and then continued through the use of periodic boundary conditions. Figure \ref{bs} illustrates\cite{ramachandran_mayavi:_2011} this for a fcc(100) model catalyst surface exhibiting two types of active sites. More complex geometries like entire nanoparticles are accessed through the thoughtful use of non-primitive unit-cells and/or pseudo reaction intermediates (e.g. declaring different active sites, effective species and kinetic parameters for every facet).

At each site an integer {\em occupation} value represents one of several possible {\em reaction intermediates} binding to this site (e.g. 1 for adsorbed {\tt O}, 2 for adsorbed {\tt CO} in the prominent CO oxidation context), including a reaction intermediate {\tt empty} (e.g. occupation value 0) and also including the possibility that a (larger) reaction intermediate extends over more than one site (in the lattice context simply achieved by additional constraints linking the occupation of neighboring sites). One specific set of occupation values on the entire lattice is called a {\em configuration} (denoted by small Latin letters $u,v,\ldots$), and a transition from one configuration to another proceeds through the occurrence of an {\em event} (denoted by small Greek letters $\alpha, \beta, \dots$). An event thus changes the occupation of one or more sites.

\paragraph{Rare-event dynamics and Markovian master equation}
The second defining characteristic is that the time evolution is characterized by a so-called rare-event  dynamics\cite{voter_extending_2002}. Due to activation barriers well exceeding thermal energies, the reaction intermediates reside most of the time in their adsorption (lattice) sites, and the events in form of the actual {\em elementary reactions} (adsorption, diffusion, reaction, desorption) happen comparably fast in between. Exploiting this separation of time scales, prevalent microkinetic theory \cite{gardiner_handbook_2004,boudart_century_2000} generally assumes that any such event occurs independent of all preceding ones, i.e. it applies a {\em Markov} approximation. The time evolution of the system (in this case the transitions from configuration to configuration through the consecutive occurrence of events) is then described by a Markovian master equation
\begin{equation}
    \label{master_eq}
    \dot{\rho}_{u}(t) = \sum_{v} \left( w_{uv}\rho_{v}(t) - w_{vu}\rho_{u}(t) \right) \quad ,
\end{equation}
where $\rho_{u}(t)$ is the probability for the system to be in configuration $u$ at time $t$, and $w_{vu}$ is the transition rate (in units of time$^{-1}$) at which configuration $u$ changes to configuration $v$.

\paragraph{Locality of elementary reactions}
The third defining characteristic of the systems mentioned initially is that changes in configuration due to an event are typically geometrically narrowly confined to as few as $\sim 1\text{-}10$ sites. Due to this locality it is possible and convenient to uniquely define any elementary reaction $a$ in terms of the local {\em educt} $E_{a}$ lattice configuration before and the local {\em product} $P_{a}$ lattice configuration after the event, as well as the concomitant rate constant $k_{a}$,
\begin{equation}
    a : E_{a} \xrightarrow{k_{a}} P_{a} \quad .
\end{equation}
Obviously, these local lattice configurations have to extend at least over all sites that actually change occupation due to the occurring elementary step. For a simple unimolecular CO adsorption step the local lattice configuration must e.g. contain the very site involved that changes its occupation from 0 ({\tt empty}) to 2 ({\tt CO}). For a dissociative adsorption step of O$_2$ the minimum local lattice configuration must in turn extend over the two neighboring sites that change their occupation from 0 ({\tt empty}) to 1 ({\tt O}), and for more complex reactions involving reaction intermediates covering multiple sites the minimum local lattice configurations span even larger lattice areas. In cases the local lattice configurations may need to include further nearby lattice sites, which do not change their actual occupation from educt to product configuration, but occupation value of which is a necessary information to determine the elementary reaction. This is prominently the case in the presence of lateral interactions. In order to properly capture such interactions the local lattice configuration needs to include all lattice sites within the interaction radius to uniquely define the local adsorbate environment. Imagine for the case of the afore mentioned unimolecular CO adsorption that this depends on whether or not a site neighboring the actual adsorption site is also occupied with CO. In this situation the local educt and product lattice configuration need to include the actual adsorption site $i$ (which changes its occupation) and the neighboring site $j$ to uniquely define two distinct elementary reactions:
\[ a_{1} : {\tt empty}@i;{\tt empty}@j \xrightarrow{k_{a_1}} {\tt CO}@i \]
and
\[ a_{2} : {\tt empty}@i;{\tt CO}@j \xrightarrow{k_{a_2}} {\tt CO}@i \quad .\]

In the presence of periodicity in the employed lattice there can be a large number of events that in fact all represent the same elementary reaction, just occurring at different lattice sites. The definition through the local lattice configurations allows to efficiently achieve this classification by first checking if local educt and product lattice configurations can be transformed into each other through a lateral lattice translation vector. Since an elementary reaction is not affected by any lattice configuration difference outside the local educt and product configuration this grouping correctly includes many events which only differ by the (non-changing) occupations outside these local configurations. To illustrate this consider again CO adsorption on an empty periodic surface featuring one type of active site. Given that adsorption into any of these sites is equivalent, adsorption events on sites $i$ and $j$ are different events in terms of the overall lattice configuration, yet they would both be grouped to the same elementary reaction by their identical local educt and product lattice configurations. Similarly, adsorption on site $i$ with another adsorbate present on site $k$ is again a different event, but falls still into the same elementary reaction class if there are no lateral interactions between sites $i$ and $k$, and $k$ is correspondingly outside the local lattice configuration. Notwithstanding, it is important to realize that identical local educt and product lattice configurations are only a necessary, but not a sufficient condition for the same elementary reaction. In the surface catalysis context, this is notably exemplified by Eley-Rideal type reaction events, where an adsorbed reaction intermediate is reacted off in a gas-phase scattering reaction. The local educt and product lattice configurations for such an event are identical to those describing a mere desorption process of the reaction intermediate. Yet, these are two distinct elementary reactions, which in the lattice framework is accounted for through two different rate constants.

\paragraph{Size and structure of the transition matrix}
The considerations about locality provide important insight into the structure of the overall transition matrix $\pmb{w}$ in Eq. (\ref{master_eq}). First, it can be decomposed into a sum of elementary reaction matrices as
\begin{equation}
    \label{group_w}
    w_{vu} = \sum_{a} w_{vu}^{a} \quad ,
\end{equation}
where
\begin{equation}
\label{defined_walpha}
    w_{vu}^{a} = \left\{\begin{array}{ll}
    k_{a}  & (u\rightarrow v) \in a\\
    0 & {\rm else}
    \end{array}\right.
\end{equation}
and $k_{a}$ is the reaction rate constant of elementary reaction $a$. The total number of different matrix elements is thus given by the number of inequivalent elementary reaction steps in the model.

Second, with respect to the structure of $\pmb{w}$ it is useful to define the set of {\em available events} ${\pmb{\sigma}}_{u}$ for any configuration $u$ as the set of all events $\alpha_{vu}$ that lead from configuration $u$ to any other configuration $v$
\begin{equation}
\label{def_sigma}
    {\pmb{\sigma}}_{u}=\{\alpha_{vu}| w_{vu} \neq 0\} \quad ,
\end{equation}
i.e. ${\pmb{\sigma}}_{u}$ is formed by the non-zero elements in the $u$th column of $\pmb{w}$. The locality of the elementary reactions implies that there are no events that connect largely differing lattice configurations. As such, ${\pmb{\sigma}}_{u}$ is much smaller than the total size of $\pmb{w}$, i.e. the transition matrix is sparse. For the later kMC efficiency discussion we further note that given an event $\alpha_{vu}$, all events in $\pmb{\sigma}_{v} \backslash \pmb{\sigma}_{u}$ are said to be {\em enabled} by $\alpha_{vu}$, while all events in $\pmb{\sigma}_{u} \backslash \pmb{\sigma}_{v}$ are said to be {\em disabled} by $\alpha_{vu}$. As any event is local and thus affects much less sites than the total number of sites in the lattice, for every event $\alpha_{vu}$ the number of enabled or disabled events is again much smaller than the number of available events in both $u$ and $v$, or
\begin{equation}
    \label{eq:diff_uv}
    |\pmb{\sigma}_{u} \cap \pmb{\sigma}_{v} | \gg | (\pmb{\sigma}_{u} \cup \pmb{\sigma}_{v})
    \backslash (\pmb{\sigma}_{u} \cap \pmb{\sigma}_{v}) |.
\end{equation}
This difference in size will become the more pronounced the larger the lattice that is to be simulated.

These insights into the structure of the transition matrix are important as the formal simplicity of the master equation (\ref{master_eq}) easily disguises the complexity in solving it in practice in the surface catalytic context. This is due to the sheer size of the space of all possible lattice configurations. To illustrate this, let us assume a simple surface system that exhibits only one type of active site per unit cell and allows for a minimum number of two different reaction intermediates at this site (again in the context of CO oxidation this could be adsorbed O and CO). Together with the possibility of an active site being empty, this yields three possible occupations of every site. In order to properly capture the ensemble characteristics of the system like the average TOF we typically need to explicitly simulate at least a surface area of $(10 \times 10)$ surface unit-cells that is then continued by periodic boundary conditions. The total number of configurations possible on this lattice is $3^{100}\approx 10^{47}$, and the $(3^{100} \times 3^{100})$ transition matrix ${\pmb{w}}$ features $(3^{100})^2\approx 10^{95}$ matrix entries $w_{vu}$. As discussed this matrix is sparse though, as there are no events that connect largely differing lattice configurations, and ${\pmb{\sigma}}_{u}$ for every configuration $u$ will be much smaller than $3^{100}$. Nevertheless, ${\pmb{w}}$ still contains a total number of non-zero elements that is too large to be stored directly. On the other hand, due to the translational symmetry of the lattice the total number of inequivalent matrix entries $w_{vu}$ is typically rather small and determined by the total number of inequivalent elementary reactions $a$ in the reaction network, cf. Eq. (\ref{defined_walpha}). For a CO oxidation model in the described simple surface system this total number can be as low as seven: dissociative adsorption of O$_2$, associative desorption of two adsorbed O, CO adsorption, CO desorption, O diffusion, CO diffusion, and CO+O reaction.

\subsection{Kinetic Monte Carlo}
It is clearly hopeless to explicitly solve such a high-dimensional master equation directly or even just aim to store the entire probability density in a lattice representation. The idea behind kinetic Monte Carlo (kMC) simulations is instead to achieve a numerical solution by generating an ensemble of trajectories of the underlying Markov process, where each trajectory propagates the system correctly from configuration to configuration in the sense that the average over the entire ensemble of trajectories yields the probability densities $\rho_{u}(t)$ of Eq. (\ref{master_eq}).\cite{chatterjee_overview_2007,voter_introduction_2007,reuter_first-principles_2012,fichthorn_theoretical_1991,bortz_new_1975,gillespie_general_1976} Analysis of any single (stochastic) kMC trajectory is correspondingly meaningless, unless a stationary system state (in the catalysis context: steady-state operation) allows to replace the ensemble average by a time average over one trajectory. The actual objective for a kMC computer algorithm (and in turn for a software package like {\tt kmos}) is therefore to generate such kMC trajectories. For this, the kMC code generally only needs to store the (evolving) occupation values on the lattice, as well as the inequivalent rate constants of all elementary reactions. On-the-fly it then generates and focuses on those transition rates $w_{vu}$ that are actually required to propagate the trajectory.

Differences between kMC solvers arise in the way how the event sequence is chosen and the concomitant way how the elapsed system time is determined. For the latter, one generally exploits that waiting times for uncorrelated events are Poisson distributed.\cite{fichthorn_theoretical_1991,bortz_new_1975,gillespie_general_1976}  This means that given a rate constant $k$ for an event, the probability that $n$ events occur in an interval $\Delta t$ is
\begin{equation}
    \label{poisson_dist}
    p_{n}(k, \Delta t) = (k\Delta t)^{n} {\rm e}^{-k\Delta t}/n! \quad .
\end{equation}
The waiting time between two events is then simply given by the case that no events occur
\[ p_{0}(k,\Delta t) = {\rm e}^{-k \Delta t}, \]
for which a suitably distributed random number can be directly computed from a uniformly distributed random
number $r\in]0,1]$\cite{h_numerical_2007} as
\begin{equation}
    \label{delta_t}
    \Delta t = \frac{-\ln(r)}{k} \quad .
\end{equation}

Lukkien {\em et al.}\cite{lukkien_efficient_1998} proposed a unified scheme consisting of three categories that classify existing kMC solvers: The first reaction method (FRM), the variable step-size method (VSSM), and the random selection method (RSM). We now briefly describe the essential features of each category, primarily to contrast the conceptual differences. The equivalence of all three approaches has also been shown by Lukkien {\em et al.}\cite{lukkien_efficient_1998}, such that a preference for one or the other emerges only out of efficiency considerations as discussed in the next section.

\begin{figure}
\centering{
\includegraphics[width=7cm]{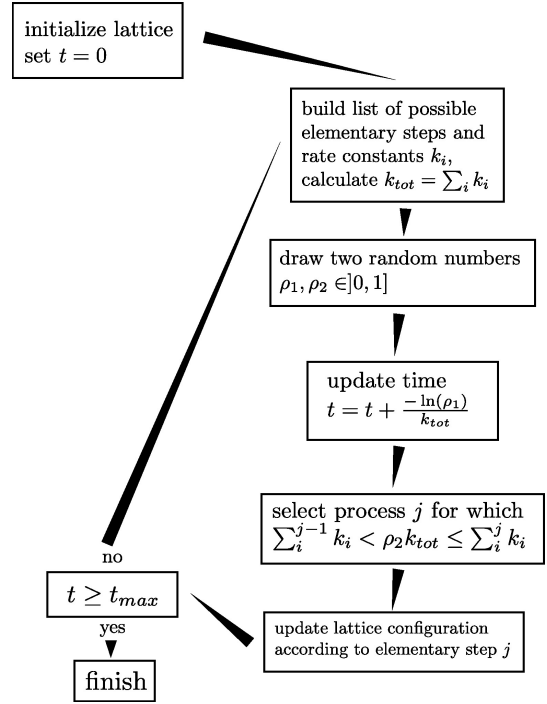}
\caption{\label{kmc_flow}The basic steps of a VSSM kinetic Monte Carlo algorithm.}}
\end{figure}

\paragraph*{FRM}
At every kMC step the FRM updates the sequence of available events $\pmb{\sigma}_{u}$ and their corresponding rate constants $\pmb{k}_{u}$. From this it calculates a sequence of time increments $\pmb{\tau} = -\ln(\pmb{r})/\pmb{k}_{u}$, where $\pmb{r}\in]0, 1]$ is a sequence of uniformly distributed random numbers. The smallest element of $\pmb{\tau}$ is selected, the elapsed time is advanced by the corresponding time increment, and the corresponding event is executed by updating the system configuration accordingly.

\paragraph*{VSSM}
At every kMC step the VSSM updates the sequence of available events $\pmb{\sigma}_{u}$ and calculates the total rate $k_{{\rm tot}, u}=\sum_{v\in \sigma_{u}} k_{vu}$. The time is advanced by $-\ln(r)/k_{{\rm tot}, u}$, where $r\in]0,1]$ is a uniformly distributed random number, and one of the available events is selected for execution with a probability weighted by its rate constant. Since VSSM is the algorithm underlying {\tt kmos}, Fig. \ref{kmc_flow} further illustrates these steps in form of a flow chart.

\paragraph*{RSM}
At every kMC step the RSM calculates $K=\sum_a k_{a}$, where the sum runs over all elementary reactions $a$. One elementary reaction is chosen with a probability weighted by its rate constant, and one of the $N_{\rm sites}$ sites in the lattice is randomly selected. The time is increased by $-\ln(r)/N_{\rm sites}K$, where $r\in]0, 1]$ is a uniformly distributed random number. The lattice is updated if the event is available at the selected site.

\subsection{Efficient Lattice kMC}
\label{kmc_efficient}

For small lattice sizes and simple kMC models containing a limited number of elementary reactions efficiency of the kMC code is generally not an issue. In particular in the context of multi-scale modeling approaches the computational costs of kMC simulations are completely negligible compared to typical costs of first-principles electronic structure theory calculations. With increasing lattice sizes and complexity of the kMC model this situation changes, in particular when considering that likely larger numbers of kMC simulations have to be run to cover different gas-phase operation conditions or when performing sensitivity analyses. In this situation, efficiency of the lattice kMC simulations becomes paramount. For the RSM algorithm a major limitation to efficiency might arise out of a diminishing probability for successful events. In the surface catalysis context this commonly arises in situations of almost fully occupied lattices and the presence of a very fast diffusion process. Such events are then predominantly chosen, but essentially never successful. For the rejection-free FRM and VSSM algorithms the main bottleneck arises instead out of the necessity to update the sequence of available events at every kMC step. A na{\"i}ve approach that is straightforward to implement (and accordingly chosen in many 'from scratch' programs) is to determine $\pmb{\sigma}_{u}$ through iteration over all lattice sites. With the number of lattice sites possibly going up to tens of thousands for entire nanoparticle simulations any retraction to such a sequential operation on the full lattice ($O(N_{\rm sites})$) will then drastically impede the overall performance.

Lukkien {\em et al.} \cite{lukkien_efficient_1998} have systematically analyzed the efficiency of the three kMC approaches and concluded on VSSM as most promising method. In line with this analysis, VSSM has also been chosen as basic algorithm underlying {\tt kmos}. In contrast to the FRM algorithm VSSM requires only two random numbers per kMC step, cf. Fig. \ref{kmc_flow}. As such its main computational burden lies in the repeating update of the set of available events and total reaction rate $k_{{\rm tot}, u}$. In contrast to the na{\"i}ve $O(N_{\rm sites})$ approach, exploitation of the locality of the elementary reactions allows to largely reduce the scaling of both these calculation steps. With respect to the set of available events this is achieved through local update procedures, thereby taking into account eq. \eqref{eq:diff_uv}. Rather than building this set anew at every kMC step, these local updates merely determine a new $\pmb{\sigma}_{v}$ from the previous set of available events $\pmb{\sigma}_{u}$ by strictly removing all disabled events ($\pmb{\sigma}_{u}\backslash \pmb{\sigma}_{v}$) and adding all enabled events (${\pmb\sigma}_{v}\backslash{\pmb\sigma}_{u}$), or formally
\[
    \pmb{\sigma}_{v}
    = ({\pmb\sigma}_{u}\backslash ({\pmb\sigma}_{u}\backslash{\pmb\sigma}_{v}))
    \cup (\pmb{\sigma}_{v} \backslash \pmb{\sigma}_{u}
    ) \quad .
\]

From the new set of available events $\pmb{\sigma}_{v}$ its corresponding total rate constant $k_{{\rm tot}, v} = \sum_{w\in \sigma_{v}} k_{wv}$ can also be calculated without iterating over the full size of the set (which would generally also scale as $O(N_{\rm sites})$). For this, not only the contained events directly, but also the number of events $N^{\rm avail}_{a,v}$ that belong to the same elementary reaction $a$ are stored. This way, if an event $\alpha_{wv}$ belonging to an elementary reaction $a$ is added to the set of available events, the corresponding counter $N^{\rm avail}_{a,v}$ is simply increased by 1, whereas if $\alpha_{wv}$ was removed, the counter is decreased. As a result one can quickly calculate $k_{{\rm tot}, v}$ as
\begin{equation}
    k_{{\rm tot}, v} = \sum_{a} k_{a} N^{\rm avail}_{a,v} \quad ,
    \label{n_avail}
\end{equation}
where the sum does not iterate over the elements in $\pmb{\sigma}_{v}$, but only over the much smaller set of elementary reactions ($O(N_{\rm react})$). Further, if the previous summation is carried out by filling an array of accumulated rates
\begin{eqnarray}
    k^{\rm acc}_{1,v} &=& 0 \\
    k^{\rm acc}_{a,v} &=& k^{\rm acc}_{a-1,v} + k_{a} N^{\rm avail}_{a,v} \quad ,
    \label{k_acc}
\end{eqnarray}
the next event $\alpha_{wv}$ can also be selected without retracting to $O(N_{\rm sites})$ operations by using a random number $r\in]0,1]$ and selecting the elementary reaction $b$ for which
\begin{equation}
    k^{\rm acc}_{b-1,v} < r k_{{\rm tot}, v} \le k^{\rm acc}_{b,v} \quad ,
\end{equation}
through a binary search ($O(\log(N_{\rm react}))$), and then selecting randomly one of the $N^{\rm avail}_{b,v}$ available events belonging to elementary reaction $b$ ($O(1)$).

As this analysis shows every required task of a VSSM lattice kMC solver can thus be carried out with a computational effort that is independent of the number of sites in the system.

\subsection{Sampling of Reaction Rates}

A central capability of kMC simulations in the context of heterogeneous catalysis is the calculation of reaction rates. Normalized to active site or surface area, corresponding TOFs yield the occurrence of any elementary reaction per time. If this elementary reaction yields a final product, then its TOF measures the overall catalytic activity with respect to this product. If there are several elementary reactions leading to different products, then the ratios of their TOFs additionally provide the selectivities.

In the context of kMC simulations a straightforward definition of the TOF per active site of any elementary reaction $a$ at any time $t$ is
\begin{equation}
\label{tof_def1}
{\rm TOF}^a(t) = \frac{\langle N^a(t) \rangle}{N_{\rm sites}} \quad ,
\end{equation}
where $N^a(t)$ is the number of times that reaction $a$ has occurred at time $t$, and the average $\langle \, \rangle$ is over a sufficiently large ensemble of kMC trajectories. Realizing that the actual occurrence of an event is given by the probability for the system to actually be in a configuration where the event is enabled times its rate constant, an equivalent definition is
\begin{equation}
\label{tof_def2}
{\rm TOF}^a(t) = \frac{\sum\limits_{u} \rho_u(t) \sum\limits_{v} w^a_{vu}}{N_{\rm sites}} \quad ,
\end{equation}
where the sums run over all configurations $u$ and $v$, $\rho_u(t)$ is the probability for the system to be in configuration $u$ at time $t$, and $w^a_{vu}$ are as defined in eq. (\ref{defined_walpha}) the transition rates of all events $\alpha_{vu}$ that correspond to the elementary reaction $a$.

In catalytic applications the primary focus is typically on steady-state operation. Even if time-dependent operation conditions and consequently time-dependent TOFs are of interest, the changes generally occur over at least mesoscopic times, and can therefore be captured through appropriate binning in constant-condition time windows. In a corresponding stationary situation the ensemble averages in eqs. (\ref{tof_def1}) and (\ref{tof_def2}) can be replaced by time averages. For the first definition this leads numerically to
\begin{equation}
{\rm TOF}^a \approx \frac{\int_0^{t_{\rm final}} N^a(t) {\rm d}t}{N_{\rm sites} \, t_{\rm final}} = \frac{N^a_{t_{\rm final}}}{N_{\rm sites}} \quad ,
\end{equation}
where $N^a_{t_{\rm final}}$ is the total number of times elementary reaction $a$ took place in a time span $t_{\rm final}$. A corresponding straightforward counting to determine TOFs is what is primarily implemented in simple 'from scratch' kMC codes. This approach becomes highly inefficient though, if small TOFs are to be measured. Due to the irregular and rare occurrence of the corresponding elementary reaction long time spans need to be simulated to sufficiently converge the TOF. In this situation it is advantageous to resort to the second TOF definition in eq. (\ref{tof_def2}),
\begin{eqnarray*}
{\rm TOF}^a &=& \frac{\sum\limits_{u} \bar{\rho}_u \sum\limits_{v} w^a_{vu}}{N_{\rm sites}} \\
&\approx& \frac{\sum\limits_{i=1}^{N_{\rm final}} \sum\limits_v w^a_{vu_i} \Delta t_i}{N_{\rm sites} \, t_{\rm final}} \\
&=& \frac{\sum\limits_{i=1}^{N_{\rm final}} k_a N^{\rm avail}_{a,u_i} \Delta t_i}{N_{\rm sites} \, t_{\rm final}} \quad ,
\end{eqnarray*}
where $N_{\rm final}$ are the number of kMC steps in the time span $t_{\rm final}$, $u_i$ is the configuration occupied at the beginning of kMC step $i$ and $\Delta t_i$ is its duration, that is the time until the simulation jumps out of $u_i$. The second equality demonstrates the efficiency of this approach, which adds to the convergence of the TOF with every kMC step even if $k_a$ is very small, and which furthermore comes at negligible overhead as $N^{\rm avail}_{a,u_i}$ is calculated at every kMC step $i$ anyway. For the determination of low TOFs this approach can therefore significantly reduce the time required for a converged sampling and is correspondingly implemented in {\tt kmos} by default.

\section{The {\tt kmos} Framework}

\begin{figure}
\centering{
\includegraphics[width=7cm]{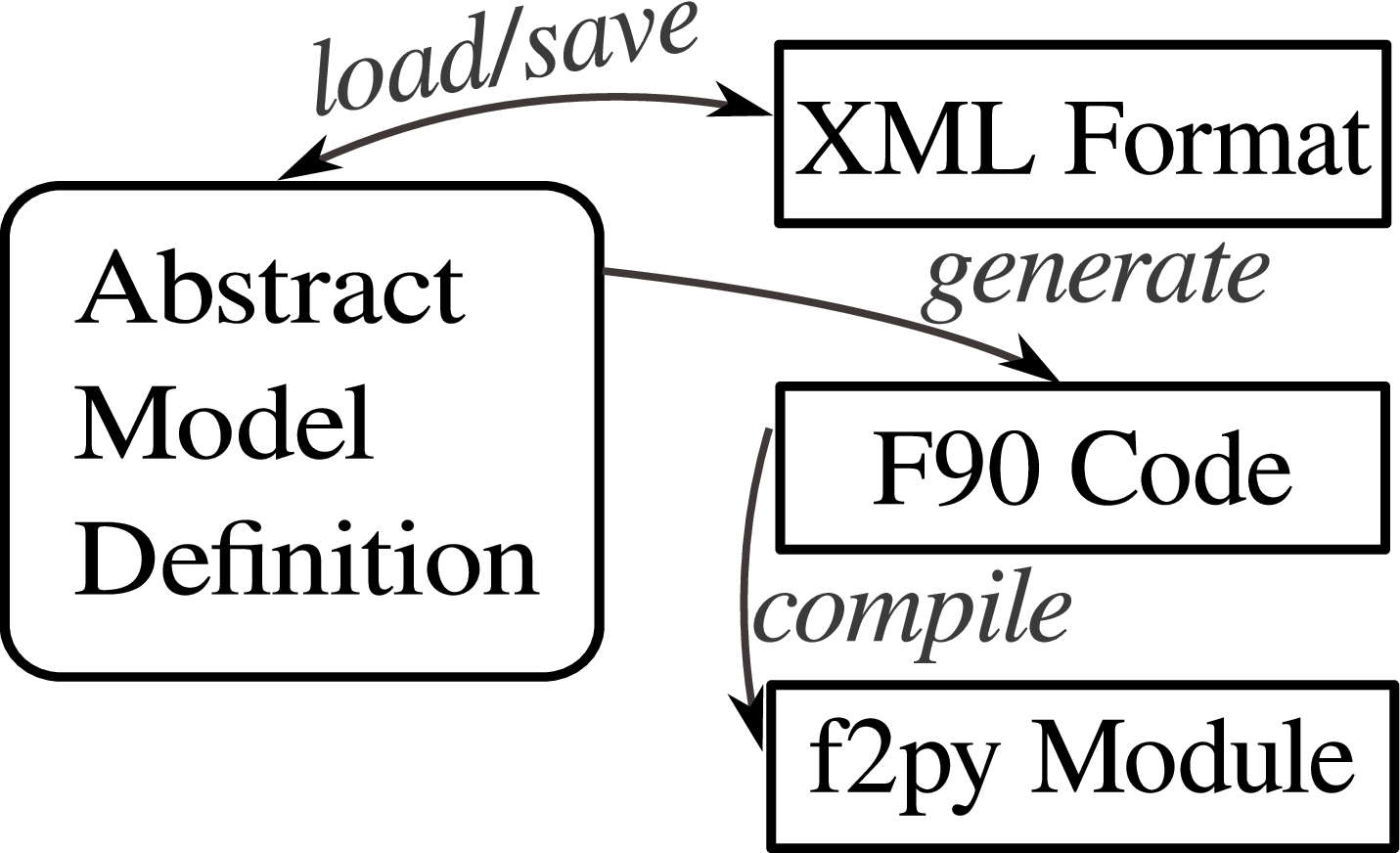}
\caption{Scheme for the flow of information in the {\tt kmos} framework. An abstract kMC model is defined by using the {\tt kmos} API. Thus, either the Python code using the {\tt kmos} API itself can serve as definitive specification of the kMC model, or the model can be stored in an XML scheme for archiving and exchange. From the abstract model definition {\tt kmos} generates tailored Fortran90 source code that performs the actual kMC simulations. This code can be compiled and exposed to Python again using f2py.}
\label{kmos_flow}}
\end{figure}

The essential idea of the {\tt kmos} approach to kMC modelling is to use a code generator to produce highly efficient code from an abstract definition of a kMC model. As further detailed below {\tt kmos} thus avoids a static and in full generality cumbersome hard-coding of the complex conditional dependencies between arbitrary events. Instead it custom tailors the code on the basis of a defined model, which in particular allows for most efficient local updates of enabled and disabled events. The general flow of information in the {\tt kmos} framework is illustrated in Fig. \ref{kmos_flow}. The following three subsections consecutively describe the three main parts apparent from this scheme: The specification of the kMC model, the code generation from the model, and how the generated code implements the VSSM kMC algorithm.

\subsection{kMC Model Definition}
\label{model_definition}

Since this part of the {\tt kmos} framework is Python based, this subsection will borrow a subset of object-oriented terminology to describe its structure: Essentially, a kMC model is a hierarchy of objects with attributes.

The information necessary to define a kMC model generally falls into two related, but distinct categories. On the one hand, there is the information required for the actual kMC simulations. This is information on the sites and lattice structure, on the reaction intermediates (code-internally called species), on general parameters like temperature or partial pressures that can be used to internally compute the rate constants, as well as all possible elementary reactions. On the other hand, there is additional information required for the analysis, in particular for an atomistic visualization of the generated kMC trajectories. This is prominently any explicit geometric information (size and shape of unit-cell, Cartesian coordinates of sites within the unit-cell, representation of substrate and reaction intermediates). In summary, this leads to the following schematic structure of the model definition:
\begin{itemize}
\item{model
    \begin{itemize}
    \item{lattice (geometry): unit cell, [sites]}
    \item{sites: name, position}
    \item{reaction intermediates (species): name, representation}
    \item{parameters: name, value}
    \item{elementary reactions: name, [conditions], [actions], rate constant}
     \end{itemize}}
\end{itemize}
Within this basic skeleton the user has to define the model specific parts. For this, one could envision some configuration file-like format. However, in particular with respect to the sequence of elementary reactions it turns out that one would have to type many very similar statements. For instance, if the same elementary reaction can be executed in several different directions due to the symmetry of the lattice. {\tt kmos} therefore offers an application programming interface (API) that allows to create each object in the model by one constructor call (in terms of object-oriented programming). This has the benefit of offering a fairly straightforward syntax, while at the same time allowing for all the flexibility and expressiveness of a high-level programming language and its control constructs such as for-loops and if-statements.

\paragraph*{Lattice definition}
\begin{figure}[ht!]
    \centering{\includegraphics[width=6cm]{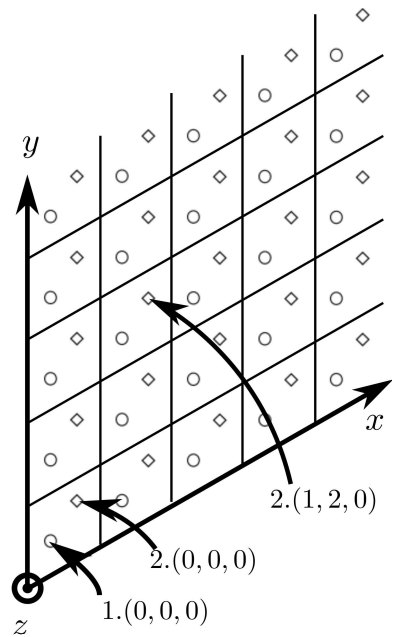}
        \caption{\label{lattice_mapping} Illustration of the lattice representation using the four-tuple {\tt n.(x, y, z)},
        where {\tt x, y, z} are the integer coordinates of the unit cell, and {\tt n} goes over the different active
        sites within the unit cell.}}
\end{figure}

The system is represented as a finite lattice with periodic boundary conditions. At present {\tt kmos} only supports one global Bravais lattice; a limitation that we intend to overcome in future work. Multiple active sites within the unit cell are accounted for through a basis. Each site is defined through a unique name. Internally every lattice point can thus be represented with a four-tuple {\tt n.(x, y, z)}, where {\tt x, y, z} are the integer coordinates of the unit cell, and {\tt n} goes over the different active sites within the unit cell as illustrated in Fig. \ref{lattice_mapping}. Naturally this scheme can describe one-, two-, or three-dimensional lattices by setting 2, 1, or 0 entries to zero respectively. By default {\tt kmos} enforces periodic boundary conditions by internally expanding the lattice by one unit cell along each lattice axis. When interested in modeling a finite lattice, this feature can be blocked by defining an inactive dummy reaction intermediate and initializing the edges of the simulated geometry with it. For visualization the shape and size of the unit cell can be specified, as well as the fractional Cartesian coordinates of all active sites within the unit cell.

\paragraph*{Reaction intermediate definition}
Reaction intermediates are specified through a unique name, and internally get assigned an integer value. A species {\tt empty} needs to be explicitly defined. Site blocking, e.g. in multidentate adsorption or to mimic infinitely repulsive lateral interactions, can be achieved through the definition of dummy species (say {\tt A$\_$blocked} as additional dummy for a reaction intermediate {\tt A} covering multiple sites). In the specification of the elementary reaction, the blocked sites are then occupied with the dummy, which thus prevents them from being {\tt empty} for other elementary reactions. Special boundary conditions such as a source or a drain that continuously inserts or removes surface intermediates at the edges of the lattice can similarly be modeled by using such special intermediates and corresponding elementary reactions. For the purpose of visualization it is possible to enter a string in the atoms-object-constructor form as understood by the Atomic Simulation Environment (ASE).\cite{bahn_object-oriented_2002}

\paragraph*{Elementary reaction definition}

\begin{figure}[ht!]
\centering{
\includegraphics[width=7cm]{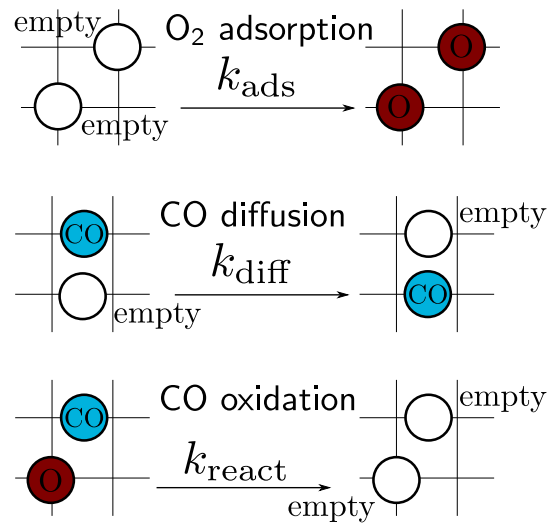}
}
\caption{\label{adsorption_es}Graphical representations illustrating the defining characteristics of elementary reactions: Dissociative O$_2$ adsorption (top panel), CO diffusion (middle panel) and CO oxidation (bottom panel).}
\end{figure}

The elementary reactions are defined in terms of the occupations in the local educt and local product lattice configuration, as well as the corresponding rate constant. Some sample definitions are depicted in Fig. \ref{adsorption_es}. For elementary reactions involving more than one site, other involved sites are specified by relative vectors in the four-tuple representation. If the central site used to define the elementary reaction is e.g. a {\tt bridge} site and another {\tt bridge} site in the unit cell in the positive direction of the first lattice vector is involved, then this additional site is referred to as {\tt bridge.(1,0,0)}. Since each element (site and occupation) of the local educt lattice configuration acts as a requirement that the elementary step can be executed, it is coined {\tt Condition}. The definition of an elementary reaction can in principle contain an arbitrary number of such {\tt Condition}s, though there are limits on the number that the compiler can process as discussed in the next section. Nevertheless, this allows to describe fairly complex elementary reactions involving lateral interactions, concerted processes, bystander adsorbates, multidentate adsorption, and even some reconstruction of the underlying lattice structure. Each element (site and occupation) of the local product lattice configuration describes a change induced by the elementary reaction and is thus coined {\tt Action}. Internally, {\tt Condition} and {\tt Action} are identical data types, but for sake of clarity different class names are used.

\paragraph*{Rate constant expressions and parameters}
{\tt kmos} accepts hard-coded values for the rate constants of the individual elementary reactions. However, in the context of surface catalysis the rate constants are often calculated using expressions such as
\[
    k = \frac{k_{\rm B}T}{h}\exp(-\beta \Delta G) \quad ,
\]
for activated surface processes or
\[
    k = \frac{p_{i}A}{\sqrt{2\pi m_{i} k_{\rm B}T}}
\]
for adsorption processes of ideal-gas particles (see e.g. ref. \cite{reuter_first-principles_2009}, which also includes the definition of the various parameters appearing in these expressions). Since it is convenient to quickly iterate external parameters (like temperature $T$ and partial pressures $p_i$) or directly change activation barriers $\Delta G$ for example in a sensitivity analysis study, {\tt kmos} also allows to directly enter such mathematical expressions for the rate constants as strings, which are later evaluated at runtime for the parameters currently present. Since these evaluations are quite expensive, they are only updated if any of the parameters change though.

\subsection{Code Generator}
\label{code_generator}

As discussed in section \ref{kmc_efficient} the main efficiency driver of a VSSM-based kMC code is the local update procedure, with its concomitant determination of disabled and enabled events. This local update procedure is also the only heavily model-dependent part of any kMC program, whereas as detailed in the next section all other parts of the actual kMC algorithm can be written in a generic way. Complicating matters, practical kMC work typically involves frequent changes of the kMC model (refinement through addition of new elementary reaction processes, consideration of further sites and reaction intermediates etc.). These changes require modifications of the code in typically as many locations as there are elementary reactions, since each new reaction might be affected by all existing elementary reactions while it can also affect every existing elementary reaction ({\em vide infra}). Doing these modifications by hand (as in the early 'from scratch' codes) is therefore not only highly cumbersome, but also extremely prone to human error. The modifications concern furthermore precisely that part of the code that determines the overall efficiency and should therefore not be implemented in an unoptimized way. {\tt kmos}' answer to this situation is to fully automatize this aspect of the work by outsourcing it to a secondary code generator program. On the basis of a defined kMC model this code generator writes the required update procedure in a compilable program language (Fortran90), which consecutively can be included in the remaining kMC program. This way, one gets the best of two worlds: A flexible high-level interface for defining kMC models and at the same time an optimized low-level implementation of the model.

The key to design an efficient code generator is to carefully reflect every logical dependence in the model to infer as many decisions as possible during the generation step and minimize the number of computational steps at runtime. This is quite different from traditional programming approaches as reflected in the following explanation, since the two levels of algorithmic description are inevitably entangled. As stated above every elementary reaction is defined in terms of {\tt Condition}s and {\tt Action}s, and in turn each of these is defined by a relative site coordinate and a concomitant species occupation. All {\tt Condition}s need to be satisfied for an event representing the elementary reaction to become enabled and only one of the {\tt Condition}s needs to be dissatisfied for an event to be disabled.

In order to implement the required updates of the set of available events $\pmb{\sigma}_{v}$ after an event $\alpha_{vu}$ has been selected, some events have to be added and some have to be removed. Removing events is conceptually and computationally simpler than adding, since removing does not require any inspection of the actual lattice configuration or evaluation if {\em all} {\tt Condition}s are satisfied. Instead it can be based on evaluating if any {\tt Action} of $\alpha_{vu}$ dissatisfies a {\tt Condition} of an available event in $\pmb{\sigma}_{u}$. One therefore iterates over the {\tt Action}s (that is configuration changes) due to $\alpha_{vu}$. For each {\tt Action} $i$, which is defined by a {\em species} and a {\em site}, one iterates over the sequence of elementary reactions.
For each elementary reaction $b$ the first check is if $b$ contains at least one {\tt Condition} $j$ on the same {\em site} as in {\tt Action} $i$. If this is the case, then such a reaction $b$ could potentially have been affected by the occurrence of event $\alpha_{vu}$. We correspondingly then also check if the {\em species} of {\tt Action} $i$ {\em does not} match with the {\em species} of {\tt Condition} $j$. If this is also the case then an event $\beta_{wv}$ corresponding to elementary reaction $b$ at the lattice site where $\alpha_{vu}$ has occurred, has become disabled. Thus compilable code is generated which removes $\beta_{wv}$ from the available events $\pmb{\sigma}_{v}$, if it was enabled in $\pmb{\sigma}_{u}$.

After the lattice configuration itself is updated (by generating corresponding compilable code to change the occupation entries) the newly enabled events can be added. Again one iterates over all {\tt Action}s of $\alpha_{vu}$. For each {\tt Action} $i$ again defined by a {\em species} and {\em site} one iterates over all elementary reactions. For each elementary reaction $b$, the first check is again if it contains at least one {\tt Condition} $j$ on the same {\em site} as {\tt Action} $i$. If in addition the {\em species} of {\tt Action} $i$ {\em does} match with the species of {\tt Condition} $j$, the corresponding event $\beta_{wv}$ of elementary reaction $b$ might have been enabled by the occurrence of event $\alpha_{vu}$. Other than in the disabling procedure we now have to iterate over all other {\tt Condition}s of $\beta_{wv}$ though, which in fact involves inspection of the occupation of all sites contained in the local educt lattice configuration of $\beta_{wv}$. Only if {\em all} {\tt Condition}s are satisfied, $\beta_{wv}$ has indeed become enabled through the occurrence of event $\alpha_{vu}$. Thus compilable code is generated which iterates over all {\tt Condition}s of event $\beta_{wv}$ at the corresponding location in the lattice and checks whether in fact the {\em species} of all {\tt Condition}s of $b$ match with the {\em species} present at the relative site. If all {\tt Condition}s are satisfied, $\beta_{wv}$ is added to the available events $\pmb{\sigma}_{v}$.

In pseudo-code the combined algorithm developed above can be concisely written as follows. Code executing before compile-time (code generation) is set in roman type, while code executed at runtime is set in {\tt monospaced} type ({\em vide infra}). Variable names are set in {\em italics}. The {\bf for} statement borrows on the Python style syntax ({\bf for} $i$ {\bf in} $x \hookrightarrow$ {\em block}), which instructs to execute {\em block} on every element of $x$  and the element will be named $i$ inside {\em block}.

\noindent
---------------------------------------------------------------\\
{\bf \# Update available events for\\
\# elementary reaction $a$\\}
\\
{\bf\#Disable events}\\
{\bf for} $i=$ ({\em species, site}) {\bf in} actions of $a$\\
\,\hspace*{.3cm}{\bf for} $b$ {\bf in} elementary reactions involving $site$\\
\,\hspace*{0.6cm}{\bf for } $j$ {\bf in } conditions of $b$\\
\,\hspace*{0.9cm}{\bf if} $i$ contradicts $j$\\
\,\hspace*{1.2cm}{\tt disable $\beta_{wv}$ if enabled}\\
\\
{\tt Update lattice configuration}\\
\\
{\bf\#Enable events}\\
{\bf for} $i=$ ({\em species, site}) {\bf in} actions of $a$\\
\,\hspace*{.3cm}{\bf for} $b$ {\bf in } elementary reactions involving $site$\\
\,\hspace*{0.6cm}{\bf for} $j$ {\bf in} conditions of $b$\\
\,\hspace*{0.9cm}{\bf if} $i$ fulfills $j$\\
\,\hspace*{1.2cm}{\tt if all conditions of $\beta_{wv}$ are met}\\
\,\hspace*{1.5cm}{\tt enable $\beta_{wv}$}\\
------------------------------------------------------------------

A crucial feature of this general update algorithm is hereby that even though it requires four nested loops, the outcome of the {\tt Condition}s checked in the outermost loops is uniquely determined by the given kMC model. Rather then evaluating these conditions during the actual runtime of the kMC simulation over and over again, the {\tt kmos} code generator evaluates them beforehand and builds the outcome directly into the generated code. The parts that need to be executed at runtime consist therefore at most of two nested loops and are by construction optimized for the defined kMC model. In terms of the generated code, the most tricky part is hereby the implementation of the check, if {\em all} {\tt Conditions} of a possibly enabled event are satisfied. Checking such interdependencies between different {\tt Condition}s typically involves many memory reads over all sites of the local educt lattice configuration and thus affects the performance. So, the question arises whether there is an optimal way how to arrange the corresponding queries. The corresponding problem of constructing an optimal binary decision tree has generally been shown to be NP-complete \cite{hyafil_constructing_1976}. In the kMC context the corresponding intricacy is given in loose terms by the fact that frequent local lattice configuration motifs, which would be the basis for an optimal construction algorithm, are unknown beforehand and precisely the outcome of a kMC simulation. Accordingly, only two heuristic approaches are presently implemented in {\tt kmos} and can be selected from the command line in the code generation step ({\tt kmos export -b<code-generator>}).

In the first approach the generated code is arranged in such a way that the average number of memory accesses are likely to be minimal for the case that every outcome is equally probable. To this end, during the code generation phase, all required read accesses are collected and sorted by decreasing frequency. All possible outcomes are grouped by the result of the most common read access and accordingly written into different conditional branches. Within each branch this process is recursively repeated. This approach shows exceptional performance at runtime for kMC models without lateral interactions and few species. Though, for more complex models involving more than three different species or considerably far-ranging lateral interactions, it often produces an exceedingly large code tree (on the order of 100MB) and accordingly long compilation time (on the order of hours).

The second approach correspondingly aims at a moderate size of the generated code and for this assumes that the primary source for the existence of multiple {\tt Condition}s is the existence of lateral interactions extending over many lattice sites. All elementary reactions are then automatically grouped into sets of identical {\tt Action}s. That is each group contains elementary reactions that are identical in the sites and species that are changed in the execution, and only differ by their {\tt Conditions} that are not changed by the elementary reaction. The rationale behind this is that models involving lateral interactions contain only a few of these sets. Within each set the present lateral interactions can be determined by as few read accesses as there are lateral interactions, since one specific lateral interaction educt excludes all others within the set. The code resulting from this approach proves to be much shorter even for models involving as much as five species and up to 40 {\tt Condition}s (on the order of few MB), and the compilation time stays typically on the order of minutes.

\subsection{Kinetic Monte Carlo Solver}
\label{kmc_solver}

The generated code is combined with other generic parts to form a VSSM lattice kMC solver that follows the general flow chart shown in Fig. \ref{kmc_flow}. To realize the efficiency considerations summarized in Section \ref{kmc_efficient} this solver operates on a well designed data structure. The base of this data structure is a bijective mapping from the four-tuple {\tt n.(x, y, z)} lattice representation to a one-dimensional representation, which simply enumerates all lattice points. This mapping can be cached in 1D and 4D arrays which makes it very efficient. Any of the frequently executed core parts are then performed on the 1D representation, and only if explicit inspection of the lattice configuration is required is the trivial inverse mapping applied. As shown below the largest arrays then have a size $(N_{\rm react} \times N_{\rm sites})$, where $N_{\rm react}$ is the total number of elementary reactions and $N_{\rm sites}$ is the total number of sites. Even for very large lattices such arrays do not represent any notable memory requirements. {\tt kmos} correspondingly uses fixed array sizes and avoids dynamic data types which would require continuous memory allocation and deallocation. On this data structure the fundamental data operations to (a) determine the next event and (b) add and delete events to and from the set of available events can be executed independent of the lattice size.

\paragraph*{Data structures}
\begin{figure*}[ht!]
    \centering{\includegraphics[width=15cm]{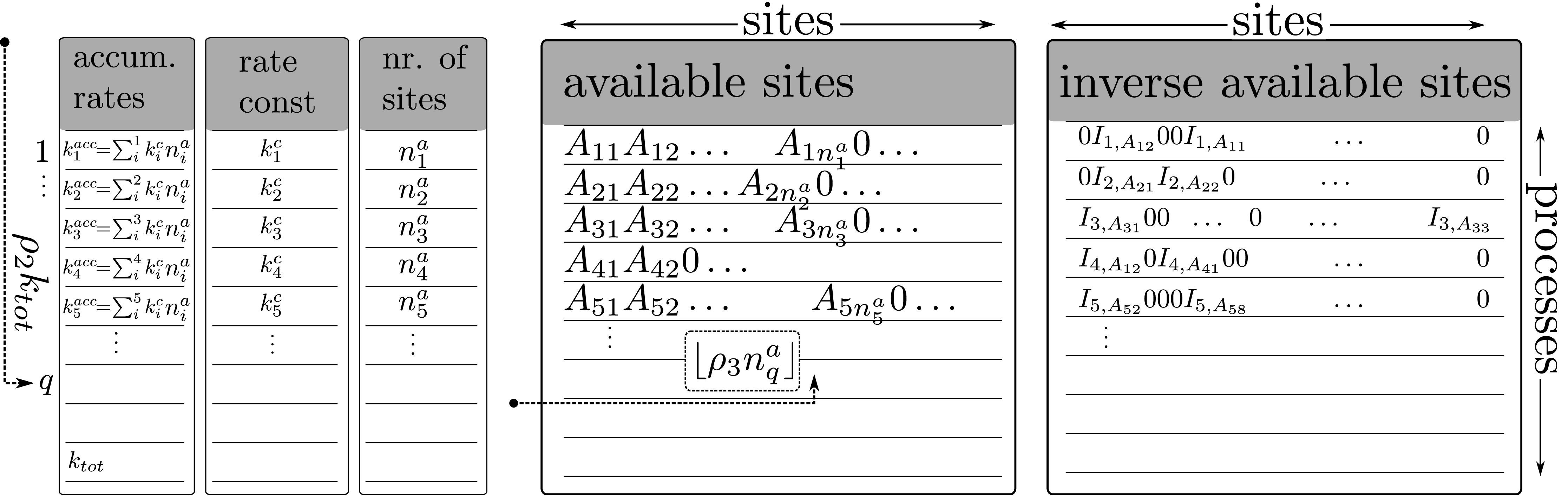}
    }
    \caption{\label{data_structures}
        Main data structures and event selection process of the {\tt kmos} VSSM-kMC solver. The array of rate constants $\pmb{k}$
        is usually unchanged during a kMC simulation. $\pmb{N^{\rm avail}}$ reflects the number of sites available for
        each elementary reaction. Using one random number $r_{2}$ an elementary reaction $a$ is selected using a binary search on the
        accumulated rate constants $\pmb{k^{\rm acc}}$. For this elementary reaction $a$ one of the available events
        is selected from ${\pmb{A}}$ using the product of a random number $r_{3}$ and the number of available sites
        $N^{\rm avail}_a$ as an index. Note that for each row $a$ in ${\pmb{A}}$ the first $N^{\rm avail}_{a}$ elements are non-zero
        after which all entries are zero. The array ${\pmb{I}}$ stores the position under which the available events are
        stored in ${\pmb{A}}$ so that all necessary updates can be executed on ${\pmb{A}}$ without traversing it.}
\end{figure*}

The deployed kMC solver operates on these 6 arrays, cf. Fig. \ref{data_structures}:
\begin{itemize}
\item{The array $\pmb{L} = \pmb{L}(N_{\rm sites})$ stores the current configuration of the system, i.e. the integer value of $L_{x}$ represents the occupation at the $x$th site.}

\item{The array $\pmb{k} = \pmb{k}(N_{\rm react})$ stores the rate constants for all elementary processes.}

\item{The array $\pmb{N^{\rm avail}} = \pmb{N^{\rm avail}}(N_{\rm react})$ stores the number of available sites for all elementary reactions, cf. Eq. (\ref{n_avail}).}

\item{The array $\pmb{k^{\rm acc}} = \pmb{k^{\rm acc}}(N_{\rm react})$ stores the accumulated rate constants, cf. Eq. (\ref{k_acc}).}

\item{The array $\pmb{A} = \pmb{A}(N_{\rm react}, N_{\rm sites})$ stores the available events. Each row of $\pmb{A}$ represents one elementary reaction and is filled from the left, i.e. an element $A_{ai} = x > 0$ tells that site $x$ is currently available for elementary reaction $a$.}

\item{The array $\pmb{I} = \pmb{I}(N_{\rm react}, N_{\rm sites})$ allows to retrieve the available events in array $\pmb{A}$. For this, if site $x$ is currently available for elementary reaction $a$ and the corresponding event is stored in element $A_{ai}$, then $I_{ax} = i$. If site $x$ is currently not available, then $I_{ax} = 0$.}
\end{itemize}

\paragraph*{Determination of the next event}
In every kMC step the solver determines the next event and therewith the concomitant elementary reaction and site as illustrated in Fig. \ref{data_structures}. First, the array of accumulated rate constants $\pmb{k^{\rm acc}}$ is updated according to the current set of available events. This includes the calculation of the total rate constant as last element $k^{\rm acc}(N_{\rm react})$. The elapsed time is updated as $-\ln(r_1)/k_{{\rm tot}}$, where $r_{1}\in]0, 1]$. Using another uniform random number $r_{2}\in]0, 1]$ and a binary search \cite{cormen_introduction_2009} an elementary reaction $a$ is determined for which $k^{\rm{acc}}_{a} < k_{\rm{tot}} r_{2} \le k^{\rm{acc}}_{a+1}$ by performing a binary search on $\pmb{k^{\rm acc}}$. Using a third uniformly distributed random number $r_{3}\in]0, 1]$ the concomitant site for the selected event is determined from array $\pmb{A}$ as the value of element $A_{ai}$, where $i = \lfloor r_{3} N^{\rm avail}_{a} \rfloor$.

\paragraph*{Update of the set of available events}
After having selected the event, that is elementary reaction and site, the code-generated part takes over to call the required additions and deletions to the set of available events, as well as the update of the lattice configuration. The prior two operations are straightforward but critical primitives of the generated local update code. In terms of the relevant data structures enabling site $x$ for elementary reaction $a$ consists of the following steps:
\begin{enumerate}
\item{Increase number of available events:\\ $N^{\rm avail}_{a} \defeq N^{\rm avail}_{a}+1$}
\item{Store site $x$: $A_{a N^{\rm avail}_{a}} \defeq x$}
\item{Assign address for site $x$: $I_{a x} \defeq N^{\rm avail}_{a} $}
\end{enumerate}
Similarly, the deletion of a disabled elementary reaction $a$ at site $x$ proceeds as:
\begin{enumerate}
\item{Overwrite site $x$ with last site enabled for $a$: $A_{a I_{a x}} \defeq A_{a N^{\rm avail}_{a}}$}
\item{Empty last site $A_{a N^{\rm avail}_{a}} \defeq 0 $}
\item{Reassign address of moved site:\\ $I_{a A_{a I_{a x}}} \defeq I_{a x}$}
\item{Empty address of deleted site: $I_{a x} \defeq 0$}
\item{Decrease available events: $N^{\rm avail}_{a} \defeq N^{\rm avail}_{a} - 1$}
\end{enumerate}

As one can see an \emph{enabling} or \emph{disabling} operation requires three or five memory transactions, respectively, and thus does not depend on the total system size or complexity. Only the search time for the next elementary reaction grows logarithmically with the number of elementary reactions $N_{\rm react}$ due to the binary search involved. However, this is not expected to become a bottleneck as this number is generally much smaller than the total number of events that have to be enabled or disabled.

This concludes all required algorithmic work in one kMC step and the next step can follow.

\paragraph*{Random numbers}
As indicated above the kMC solver requires three uniformly distributed random numbers per kMC step. {\tt kmos} relies on the pseudo random number generator (PRNG) provided by the Fortran compiler. Sometimes kMC practitioners are concerned whether such a source of randomness introduces non-physical bias to the generated kMC trajectory. We have not observed any such bias in a kMC simulation so far. In case doubt arises this can be easily tested by changing the PRNG seed conveniently in the configuration file of the compiled kMC model. Furthermore, the currently specified PRNG periods of the most commonly used compilers typically exceed the maximum number of kMC steps during one
simulation by several orders of magnitude.

\paragraph*{Overall code layout}
Having specified the kMC solver independently of any lattice geometry or specifics of elementary processes means one can reuse this part of the algorithm for virtually any lattice kMC model. This is also reflected in the structure of the overall Fortran90 code: It is subdivided into the modules {\tt base}, {\tt lattice}, and {\tt proclist}, of which {\tt base} contains the model-independent parts of the VSSM loop that has been described in this Subsection. The module {\tt lattice} replicates the {\tt base} API in terms of lattice coordinates and implements corresponding information about the model (number of lattice dimensions, numbers of sites per unit cell, and names of sites) for visualization, as well as the central VSSM loop. The third module {\tt proclist} is the one produced by the code-generator and implements how the set of available events is updated after an event has been selected in a kMC step, cf. Section \ref{code_generator}.

\subsection{Simulation Frontend}

The complete kMC model is stored in an XML text file by using the elementtree XML library. This also allows for easy archiving and exchange of models. A basic graphical user interface (GUI) is provided to visually inspect all aspects of the model definition including the elementary reactions. The generated Fortran90 code is compiled and exposed as a Python module with the f2py \cite{peterson_f2py:_2009} interface generator. {\tt kmos} offers a concise API which allows to control all runtime aspects of a compiled model including setup and evaluation, as a script or interactively using IPython \cite{perez_ipython:_2007} and numpy \cite{oliphant_python_2007}, as well as a GUI which visualizes the model geometry using ASE \cite{bahn_object-oriented_2002} and coverages and turnover frequencies using matplotlib \cite{hunter_matplotlib:_2007}, while allowing to visually change parameters during the simulation.

\section{Performance and Scaling in Practice}

\subsection{ZGB Model}

\begin{figure}[ht!]
    \centering{\includegraphics[width=7cm]{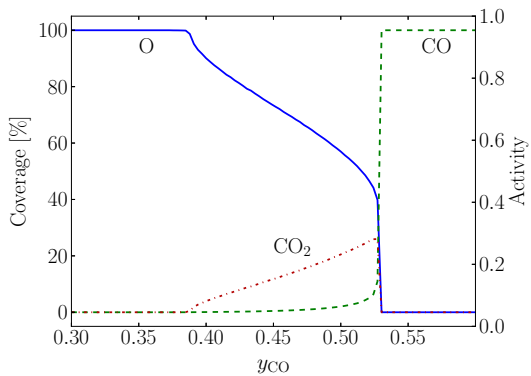}
    }
    \caption{\label{zgb_model} Coverage dependence and catalytic activity of the ZGB model as implemented using {\tt kmos}. In the idealized ZGB model catalytic activity is defined as the number of CO$_2$ molecules produced per reactant impingement.\cite{ziff_kinetic_1986}}
\end{figure}

We demonstrate the performance and scaling behavior of {\tt kmos} using a range of kMC models, and start with the seminal model by Ziff, Gulari, and Barshad (ZGB) \cite{ziff_kinetic_1986}, that has evolved into an influential reference for the development of stochastic approaches to surface catalytic processes. The original ZGB model generically considers CO oxidation at a simple cubic lattice, featuring one active site and only three elementary reactions: irreversible unimolecular adsorption of CO with rate constant $y_{\rm CO}$, irreversible dissociative adsorption of O$_2$ at two neighboring sites with rate constant $1-y_{\rm CO}$, and instantaneous CO oxidation reaction of directly neighboring adsorbed CO and O. The only free parameter of the model is thus $y_{\rm CO}$, which is varied in the range [0,1] a.u. In the context of numerical kMC simulations we realize this model by approximating the instantaneous CO oxidation reaction with an exceeding rate constant of $10^{15}$ a.u., and adding unimolecular CO and associative oxygen desorption reactions with negligible rate constants of $10^{-13}$ a.u. to mimick the irreversible adsorption. Especially the latter is necessary to prevent the system from getting trapped in completely oxygen or CO poisoned configurations, but we validated that neither the obtained results nor runtime performance depends on the particular choice of the finite rate constants chosen for these processes. Figure \ref{zgb_model} shows the resulting lattice occupations and CO$_2$ TOF in the relevant range of $y_{\rm CO}$, perfectly reproducing the two critical $y_{\rm CO}$ values of $y_{1}=0.389$ and $y_{2} = 0.527$ that delimit the O and CO coexistence at the surface and the concomitant catalytic activity.\cite{ziff_kinetic_1986} The simulations were performed on a lattice containing $(200 \times 200)$ sites, and for this benchmark system {\tt kmos} executed 2.15 million kMC steps per second on a 3.4 GHz Intel Core i7 processor with 16GB RAM. Given that the simulated elapsed time per kMC step varies with every configuration, the corresponding CPU time per million kMC steps (0.47 sec) is the only transferable benchmark property across implementations and somewhat even across models of similar complexity ({\em vide infra}).

\subsection{Literature First-Principles kMC Models}

\begin{figure}
    \includegraphics[width=8cm]{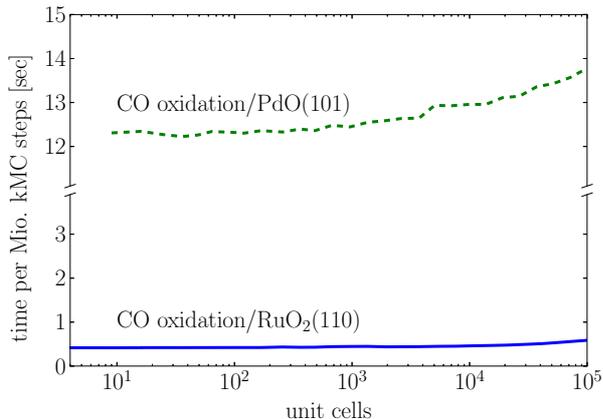}
    \caption{\label{bm_nolatint} Single-core CPU times required to execute one million kMC steps for the CO oxidation at RuO$_2$(110)
    model (solid line) and the CO oxidation at PdO(101) film model (dashed line) as a function of the simulated lattice size
    (in numbers of unit cells). For both models the {\tt kmos} performance is essentially independent of the lattice size
    in the size range relevant for catalytic applications. The simulation time is instead primarily determined by the model
    complexity. The benchmarks were carried out on a 3.4 GHz Intel Core i7 processor with 16GB RAM.}
\end{figure}

As representative examples for modern first-principles based kMC models we consider the CO oxidation model at RuO$_2$(110) as put forward by Reuter and Scheffler \cite{reuter_steady_2004,reuter_first-principles_2006} and the CO oxidation model at a thin PdO(101) film on top of Pd(100) as put forward by Rogal, Reuter and Scheffler \cite{rogal_first-principles_2007,rogal_co_2008}. The prior model does not include lateral interactions, while the latter model does include pairwise nearest-neighbor lateral interactions at an otherwise comparable number of inequivalent elementary reactions. The comparison of the two models therefore provides first insight into the performance dependence of {\tt kmos} on the number of {\tt Condition}s. Specifically, the CO oxidation at RuO$_2$(110) model includes two different active sites per surface unit cell, and a total of 26 inequivalent elementary processes (unimolecular CO adsorption and desorption, dissociative adsorption and associative desorption of O$_2$, CO and O diffusion, as well as CO oxidation and CO$_2$ decomposition).\cite{reuter_steady_2004,reuter_first-principles_2006} The PdO(101) model includes the same types of elementary reactions and also two different active sites per unit cell. In addition, it accounts for nearest-neighbor lateral interactions that modify the rate constants of all diffusion, desorption and reaction steps.

Figure \ref{bm_nolatint} shows the CPU time required to execute 1 million kMC steps for both models, again calculated on the 3.4 GHz Intel Core i7 with 16GB RAM benchmark system. Summarized is the scaling up to a maximum system size comprising $10^5$ lattice sites, which is already much larger than the $10^2-10^3$ lattice sites on which these models were reliably evaluated in the original publications. In both cases the runtime is practically independent of the lattice size, confirming the scaling considerations made in Section \ref{kmc_efficient}. The moderate increase is presumably due to a less efficient utilization of the processor cache. Memory limitations eventually also determine the maximum system sizes that {\tt kmos} can currently handle (outside the size range shown). The runtime is instead critically determined by the system complexity, and in particular by the number of {\tt Condition}s implied by the model. Even though the RuO$_2$(110) model contains a larger number of elementary reactions than the ZGB model, the CPU time per million kMC steps is thus almost the same (0.5 sec). In contrast, the pairwise lateral interactions in the PdO(101) model and the concomitant number of {\tt Condition}s increase this CPU time by a factor of $\sim 25$.

\subsection{Random Models}

To further investigate the performance dependence on the model complexity we finally consider random models with varying number of active sites per unit cell, number of possible reaction intermediates (species), number of {\tt Condition}s per elementary reaction, and number of elementary reactions using the moderate-code-size generator. That is, first $N_{\rm sites}$ sites are initialized ({\tt site1}, {\tt site2}, \dots). Next, $N_{\rm species}$ are initialized ({\tt species1}, {\tt species2}, ..). Using these ingredients we construct $N_{\rm react}$ times a pair of elementary reactions: A forward reaction which consists of $N_{\tt Condition}$ {\tt Condition}s with the default species {\tt empty} on $N_{\tt Condition}$s random sites within a finite cut-off radius and corresponding {\tt Actions} on these same sites with random species. The corresponding backward reaction uses the {\tt Actions} of the forward reaction as {\tt Condition}s and uses {\tt empty} on these same sites as {\tt Condition}s. By creating all elementary reactions in such pairs we automatically prevent dead-lock configurations in which no events are available. All elementary reactions have the same constant rate constant, and in all cases, the simulated lattice size was $(20\times 20)$ unit cells, as the preceding sections have shown that the performance scaling with model complexity is independent of the system size.

\begin{figure*}[ht!]
    \centering{
        \includegraphics[width=7cm]{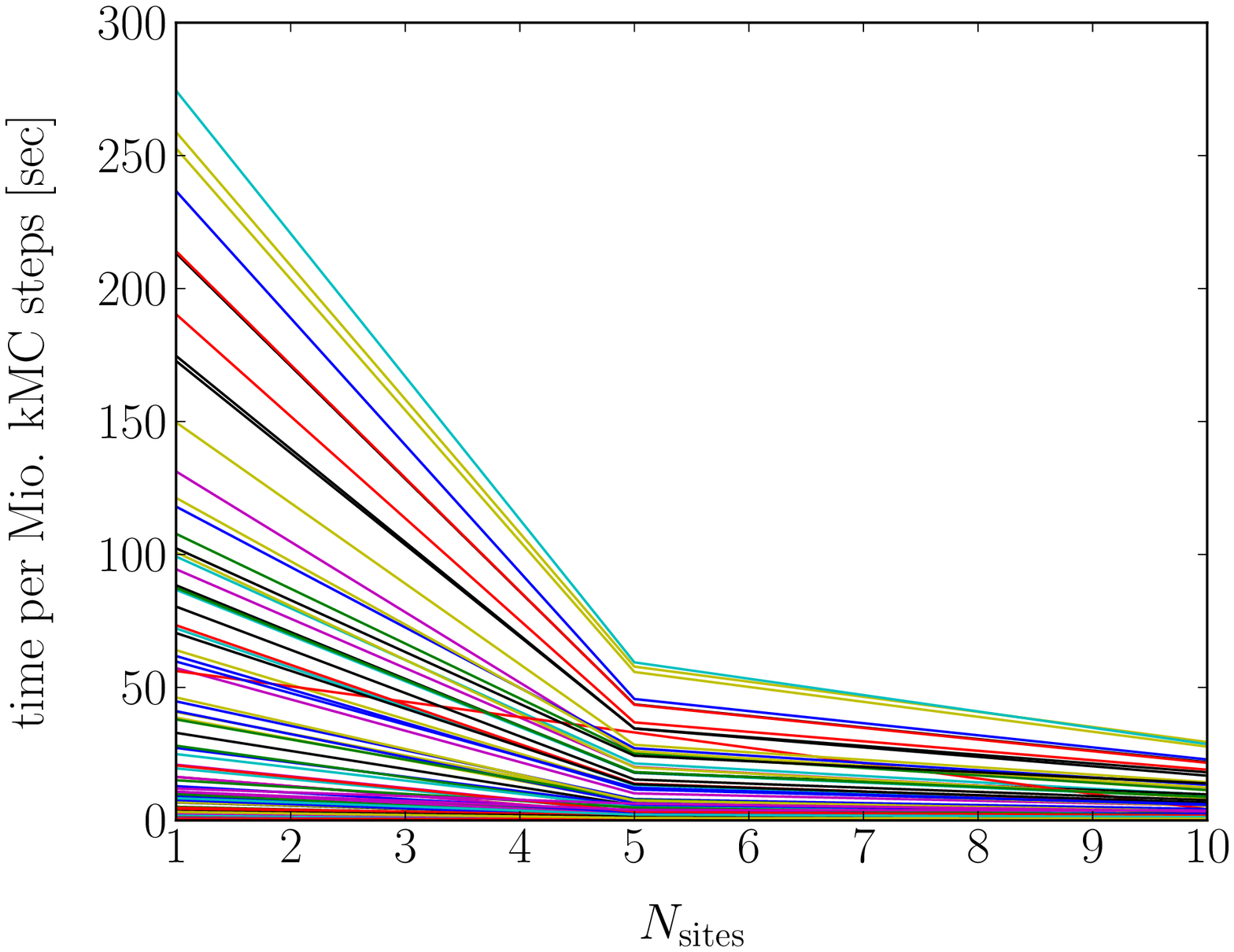}
        \includegraphics[width=7cm]{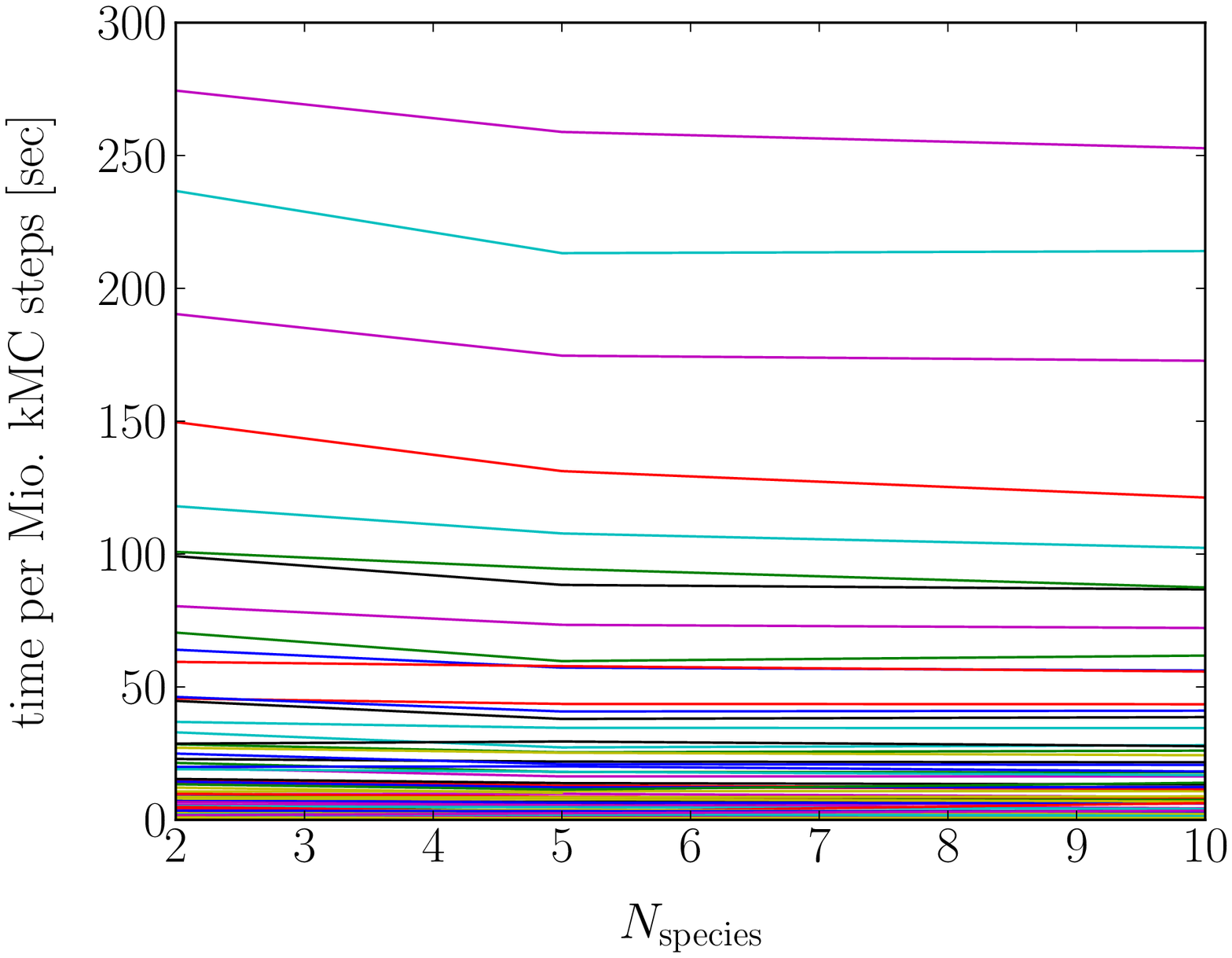}
        \includegraphics[width=7cm]{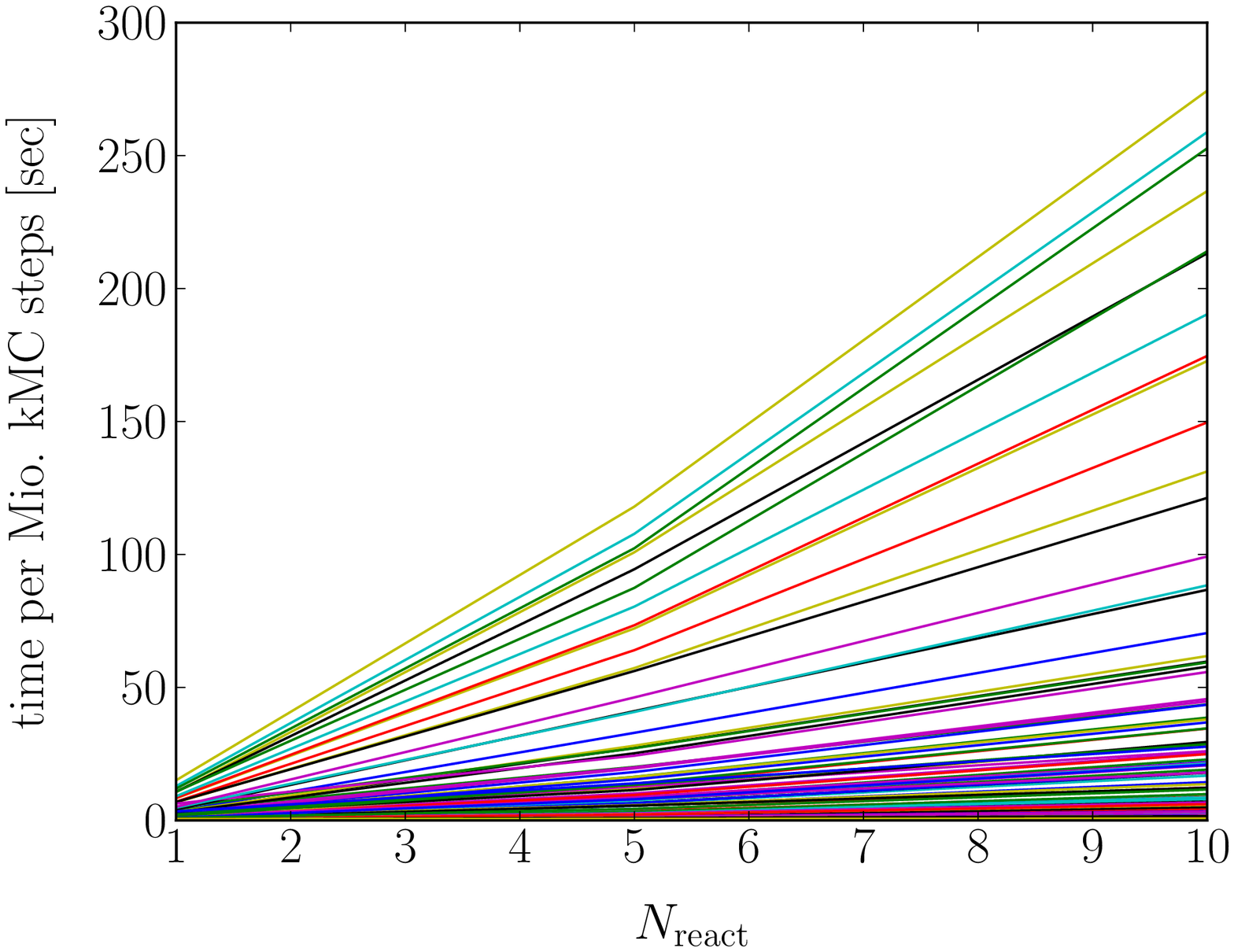}
        \includegraphics[width=7cm]{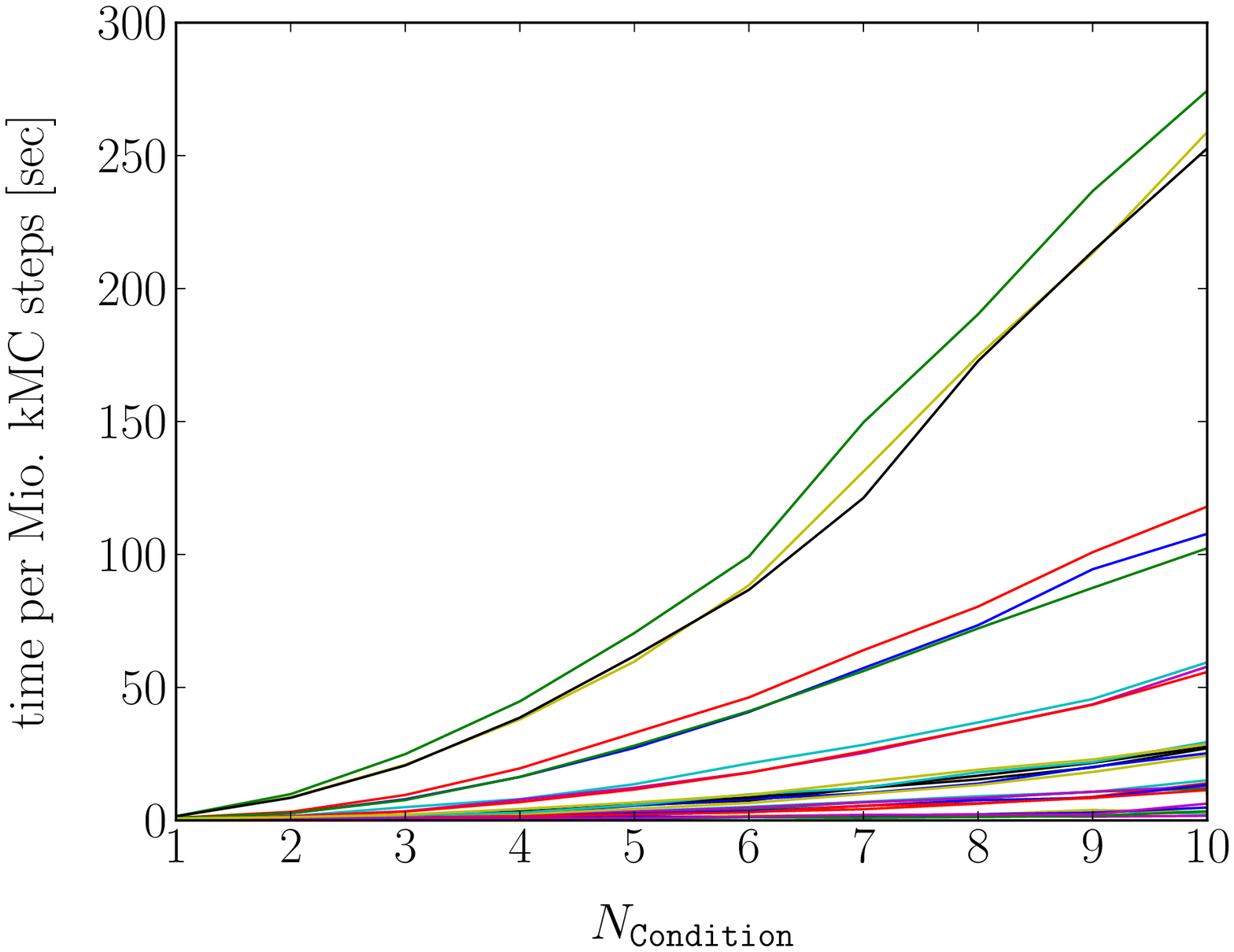}
        }
    \caption{\label{bm_times} Single-core CPU times required to execute one million kMC steps for various random models on the
    standard 3.4 GHz Intel Core i7 benchmark processor with 16GB RAM. Each panel shows the dependence along one model parameter
    ($N_{\rm sites}, N_{\rm species}, N_{\tt Condition}$ and $N_{\rm react}$). Continuous lines connect simulation results
    obtained for random models in which all other parameters are identical, i.e. in the $N_{\rm react}$ panel each line represents
    the runtime dependence on $N_{\rm react}$ for a constant set of $N_{\rm sites}, N_{\rm species}$ and $N_{\tt Condition}$.}
\end{figure*}

Using each combination of $N_{\rm sites}\in [1, 5, 10]$, $N_{\rm species}\in [2, 5, 10]$, $N_{\tt Condition}\in[1,2,\dots,10]$, and $N_{\rm react}\in[1, 5, 10]$, we evaluate the single CPU time to execute 1 million kMC steps as in the preceding subsections. Figure \ref{bm_times} compiles the obtained results, i.e. the dependence on each model dimension. To further analyze the obtained dependencies the obtained runtimes are fitted to
\[
t \propto (N_{\rm sites})^{a} \times (N_{\rm species})^{b} \times (N_{\rm react})^{c} \times (N_{\tt Condition})^{d} \quad ,
\]
yielding $a\approx-0.99$, $b\approx -0.07$, $c\approx1.24$, and $d\approx 2.00$. This shows empirically that the runtime depends approximately quadratically on the number of {\tt Condition}s per elementary step. Furthermore it demonstrates that the runtime is
basically independent of $N_{\rm species}$ and slightly above linear with $N_{\rm react}$, confirming the observations made above with the first-principles kMC models. Last, it reveals the seemingly paradoxical result that the runtime decreases with $N_{\rm sites}$. This can be rationalized by the fact that for a fixed number of elementary reactions $N_{\rm react}$ the probability that different events enable or disable each other shrink with increasing content in the unit-cell. This leads on average to fewer add or delete operations to the set of available events and concomitantly to decreasing runtimes. We stress though that this dependence is of little relevance for physically motivated kMC models, since there the number of elementary reactions $N_{\rm react}$ is expected to grow at least linearly with the number of different active sites $N_{\rm sites}$. In practice, kMC models will also exhibit a different number of {\tt Condition}s for each elementary reaction. As such, the benchmark results obtained for the random models should not be taken too literally. Nevertheless, they should convey a useful rough orientation for the to-be-expected runtimes of real kMC models featuring corresponding numbers of sites, species, and elementary reactions, as well as average number of {\tt Condition}s per reaction.

Most centrally, the results obtained with the random models underscore that the number of {\tt Condition}s is the most critical property in terms of runtime. This is not critical for model complexities currently addressed, in particular in the context of first-principles kMC simulations of surface catalysis. Notwithstanding, if eventually more than 5-6 reaction intermediates over multiple active sites and with extensive lateral interactions need to be handled, this will change -- and the current moderate\_code\_size code generating algorithm might also reach the capabilities of current compilers. Long-term systematic improvements of {\tt kmos} and its efficiency are therefore best spent on this aspect and in particular the binary decision tree to group the queries checking on the interdependencies between different {\tt Condition}s.

\section{Summary}

We have presented the open source \cite{_gnu.org_2007} package {\tt kmos}, which offers a versatile software framework for efficient lattice kMC simulations, in particular in surface catalysis. {\tt kmos} can handle site-specific reaction networks of arbitrary complexity in one- to three-dimensional lattice systems, involving multiple active sites in periodic or aperiodic arrangements, as well as site-resolved pairwise and higher-order lateral interactions. For the kMC model definition {\tt kmos} offers an extended application programming interface. On the basis of this model definition, a code generator creates an optimized low-level implementation of the main efficiency driver of a VSSM-based kMC code, the local update procedure that determines the disabled and enabled events after the execution of each kMC step. Together with a well designed data structure, this leads to an efficient kMC solver the runtime performance of which is essentially independent of the lattice size. Instead, the runtime sensitively depends on the model complexity and there in particular on the number of {\tt Condition}s implied by the elementary reactions. For the complexity of reaction networks currently perceivable in the surface catalytic context this is not critical. Should higher efficiency eventually be required, improvements to this end and the code generation algorithm either through improved binary decision trees or parallelization strategies could become of interest. 

Next to the efficiency, {\tt kmos} other core objective is a most user-friendly implementation, execution, and evaluation of lattice kMC simulations. For this the API allows to control all runtime aspects interactively, through scripts or via a basic graphical user interface. Enhancing the reproducibility and reusability of the kMC models through a standard file format, {\tt kmos} is thus hoped to contribute to a further, wide-spread use of the kMC approach by an extending user community. 

\section*{Acknowledgements}
MJH would like to thank J{\"o}rg Meyer and Sergey Levchenko for stimulating discussions. We gratefully acknowledge support from the German Research Council (DFG).


\appendix
\bibliographystyle{elsarticle-num}

\end{document}